\documentclass[]{iopart}

\expandafter\let\csname equation*\endcsname\relax
\expandafter\let\csname endequation*\endcsname\relax

\usepackage[unicode]{hyperref}
\usepackage[all]{hypcap}
\usepackage{amssymb,amsmath,verbatim,mathtools,needspace,enumitem,etoolbox,graphicx,microtype,pifont,url}
\usepackage[usenames,dvipsnames]{xcolor}
\usepackage[T1]{fontenc}
\usepackage[utf8]{inputenc}
\usepackage[normalem]{ulem}
\definecolor{linkcolor}{rgb}{0.0,0.3,0.5}
\hypersetup{colorlinks=true,linkcolor=linkcolor,citecolor=linkcolor,filecolor=linkcolor,urlcolor=linkcolor}
\usepackage{orcidlink}

\newcommand{\ssim}{\mathchar"5218\relax\,} 

\newcommand{\Mirr}{\ensuremath{M_{\rm irr}}}
\newcommand{\Mspin}{\ensuremath{M_{\rm spin}}}

\newcommand{\bham}{{School of Physics and Astronomy \& Institute for Gravitational Wave Astronomy, University of Birmingham, Birmingham, B15 2TT, UK}}
\newcommand{\milan}{{Dipartimento di Fisica ``G. Occhialini'', Universit\'a degli Studi di Milano-Bicocca, Piazza della Scienza 3, 20126 Milano, Italy}}
\newcommand{\infn}{{INFN, Sezione di Milano-Bicocca, Piazza della Scienza 3, 20126 Milano, Italy}}
\newcommand{\cambridge}{{Department of Applied Mathematics and Theoretical Physics, Centre for Mathematical Sciences, University of Cambridge, Wilberforce Road, Cambridge CB3 0WA, UK}}
\newcommand{\caltech}{Theoretical Astrophysics 350-17, California Institute of Technology, 1200 E California Boulevard, Pasadena, CA 91125, USA}
\newcommand{\jhu}{Department of Physics and Astronomy, Johns Hopkins University, 3400 N. Charles Street, Baltimore, Maryland 21218, USA}

\begin{document}

\begin{center}
\title[D.~Gerosa et al.]{The irreducible mass and the horizon area of LIGO's black holes}
\end{center}

\author{
Davide Gerosa$^{1,2,3}$ \orcidlink{0000-0002-0933-3579},
Cecilia Maria Fabbri$^{1}$ \orcidlink{0000-0001-9453-4836},\\
Ulrich Sperhake$^{4,5,6}$ \orcidlink{0000-0002-3134-7088}
}
\vspace{0.1cm}
\address{$^{1}$~\milan}
\address{$^{2}$~\infn}
\address{$^{3}$~\bham}
\address{$^{4}$~\cambridge}
\address{$^{5}$~\caltech}
\address{$^{6}$~\jhu}

\ead{\href{mailto:davide.gerosa@unimib.it}{\rm davide.gerosa@unimib.it}}

\setcounter{footnote}{0}

\begin{abstract}
The mass of a Kerr black hole can be separated into irreducible and rotational components ---the former is a lower limit to the energy that cannot be possibly extracted from the event horizon and is related to its area. Here we compute the irreducible masses of the stellar-mass black holes observed by gravitational-wave interferometers LIGO and Virgo. Using single-event data, we present a re-parametrization of the posterior distribution that explicitly highlights the irreducible and rotational contributions to the total energy. We exploit the area law to rank the black-hole mergers observed to date according to their irreversibility, thus providing a guide to selecting events for targeted tests of General Relativity. Using population fits, we compute the rate by which the total area of black-hole horizons increases due to the observable mergers.
\end{abstract}

\section{Some energy is lost inside black holes}

Penrose \cite{1969NCimR...1..252P} first showed that mass can be extracted from a Kerr black hole (BH). In astrophysics, this same process is believed to power quasars at high redshifts~\cite{1977MNRAS.179..433B}. Crucially, extracting energy from a BH comes at the price of also extracting angular momentum. It turns out that angular momentum runs out first, such that a repeated series of Penrose processes leaves behind a lighter, non-spinning BH.  Christodoulou~\cite{1970PhRvL..25.1596C} soon realized that the mass of any Kerr BH must have an ``irreducible'' component which cannot be dissipated even with extraction episodes that are maximally efficient. The irreducible mass is thus an absolute lower limit to the energy that is inexorably stored inside the horizon and cannot be possibly be taken out (at least excluding quantum effects).

Consider a Kerr BH of mass $M$ and angular momentum $S=M^2\chi$, where $\chi\in [0,1]$. The irreducible mass is given by 
\begin{equation}
\Mirr = %
 M \sqrt{\frac{1+\sqrt{1-\chi^2}}{2}}
\label{mirrdef}
\end{equation}
such that $\Mirr = M$ for $\chi=0$ and $\Mirr = M/\sqrt{2}$ for $\chi=1$. This is equivalent to writing %
\begin{equation}
M^2 = \Mirr^2 + \Mspin^2\,, \label{mtotcontr}
\end{equation}
with 
\begin{equation}
\Mspin = \frac{S}{2 \Mirr} =   M \sqrt{\frac{1-\sqrt{1-\chi^2}}{2}}\,.
\label{mspindef}
\end{equation}
The mass $M$ can thus be decomposed into an irreducible contribution $\Mirr$ and a rotational contribution $\Mspin$. Note that these pieces do not add linearly but rather, in the words of Misner, Thorne, and Wheeler \cite{1973grav.book.....M}, %
\emph{``combine in a way analogous to the way rest mass and linear momentum combine to give energy, $E^2=m^2+p^2$}.'' %

The irreducible mass is related to the area of the 
BH horizon (or more precisely, its two-dimensional section)  %
by %
\begin{equation}
A = 16 \pi M_{\rm irr}^2\,.
\end{equation}
The realization that the mass of a BH has an irreducible component is a manifestation of  Hawking's area theorem \cite{1971PhRvL..26.1344H}, which states that the variation of the total horizon area of BHs in the spacetime has to be non-negative. The BH horizon area thus acts much like entropy in thermodynamics ---an analogy that has far reaching consequences in theoretical physics~\cite{2001LRR.....4....6W}. 

These fundamental considerations on the nature of BHs apply  to the systems that are now observed in gravitational waves (GWs)\footnote{It is interesting to note that Hawking's first proof of the area law \cite{1971PhRvL..26.1344H} quotes Weber's claimed GW detection \cite{1969PhRvL..22.1320W} as key motivation to pursue the calculation.}  \cite{2016PhRvL.116f1102A}. The ground-based interferometers LIGO and Virgo are sensitive to the mergers of BHs of masses $\lesssim 100 M_\odot$ at redshifts $\lesssim 1.5$ and have so far detected dozens of such events \cite{2019PhRvX...9c1040A, 2021PhRvX..11b1053A,2021arXiv211103606T}. 
The goal of this paper is to characterize the irreducible contribution to the masses of the BHs observed in GW astronomy, thus inferring the amount of energy that is forever lost inside those BH horizons. Our estimate relies on LIGO/Virgo data and is thus restricted to merging BH binaries that emit GWs in the sensitivity band of the detectors. The 
 larger population of BHs that are not in binaries and/or have very different masses is not captured by our investigation. 

Reference~\cite{2020arXiv200512201D} has reported measurements of the irreducible masses from samples of BH candidates identified through their electromagnetic emission, including both supermassive BHs in active galactic nuclei and stellar-mass BHs in x-ray binaries. Our analysis explicitly exploits the Bayesian characterization of the available GW data
at both  single-event and population levels. Furthermore, the binary nature of the GW systems (compared to the single BHs observed electromagnetically) allows us to quantify the irreversibility of the underlying merger process. Other applications of the irreducible mass in GW astronomy include Ref.~\cite{2017PhRvL.119y1103V}, where some LIGO events have been analyzed using a Bayesian prior that is flat in rotational energy. GW data have been used to experimentally confirm the  validity of the area law~\cite{2018PhRvD..97l4069C,2021PhRvL.127a1103I}, which  requires separating the signal into pre- and post-merger chunks and then checking that the sum of the areas of the component BHs is smaller than that of the  remnant. Reversing the argument, one can also assume that the area law holds and use it to constrain some of the BH-binary parameters \cite{2021arXiv211206856H}. The irreducible mass is also routinely computed in numerical relativity to characterize the properties of simulated spacetimes \cite{2010nure.book.....B} while here we compute it for the observed systems.

In this study, we concentrate on the \emph{characterization} of the available GW data in light of some specific aspects of the underlying theory, namely the existence of an irreducible component to the BH mass and the area law. Where possible, we interpret our findings using 
toy models and back-of-the-envelope estimates. %
Test of the theory and development of new predictions are left to future analyses. We present a suite of complementary explorations, all focused on highlighting a so-far-unexplored description of the observed systems.

We begin our presentation with a semi-analytical model of the expected posterior distribution of the irreducible mass and the horizon area (Sec.~\ref{simplemodel}). This is useful to build intuition on quantities that are not often considered in observational GW astronomy. We then post-process   data from individual GW events to illustrate the current constraints on the irreducible and rotational contributions to the observed masses (Sec.~\ref{individual}). This allows us to rank the detected events based on the ``irreversibility'' (i.e. the increase of entropy) of the underlying merger process and locate, among the numerous systems detected so far, some promising candidates for future tests of gravity  (Sec.~\ref{increasearea}). Using the entirety of the GW sources detected so far, we formulate (for the first time to the best of our knowledge) and evaluate a population version of Hawking's area law (Sec.~\ref{popsec}). This allows us to place a bound on the overall energy rate that is lost inside event horizons because of the BH mergers. We conclude by clarifying how our findings can be summarized into a re-parametrization of the BH properties (Sec.~\ref{concl}). In particular, we stress that computing quantities such as the irreducible mass and the horizon area is a straightforward operation that could be implemented in future GW-catalog reports at a negligible cost. We use natural units where $c=G=1$ but reinstate those constants in a few instances for clarity.

\section{A simple model}
\label{simplemodel}

First, we wish to develop some intuition on the expected  posterior distribution of the irreducible mass and the horizon area using a  simplified model where one only measures the mass of the BH and not its spin.

Let us define 
\begin{equation}
f \equiv \frac{\Mirr}{M} =  \sqrt{\frac{1+\sqrt{1-\chi^2}}{2}} \in \left[\frac{1}{\sqrt{2}},1\right]\,.
\end{equation}
We assume that the spin magnitude $\chi$ is measured to  be  uniformly distributed in $[0,1]$, which is equivalent to the prior distribution used in the vast majority of GW analyses (i.e.~the data are uninformative).
In this case, the probability density function of $f$ is 
\begin{equation}
\pi(f) = 2\frac{2 f^2-1}{\sqrt{1 - f^2}}\,.
\end{equation}
Let us also assume that the measurement of the BH mass $M$ follows a Gaussian distribution $\mathcal{N}(\mu,\sigma)$ with mean $\mu$ and width %
$\sigma$.  The distribution of the irreducible mass can thus be written down semi-analytically as
\begin{equation}
p(\Mirr)  = \frac{\sqrt{2/\pi}}{\sigma} \int_{1/\sqrt{2}}^1 \exp\left[{-\frac{(\Mirr/f -\mu)^2}{2\sigma^2}}\right]  \frac{2 f^2-1}{f \sqrt{1 - f^2}}  \,\, {\rm d}f\,. \label{pmirr}
\end{equation}
Note that $\mu$ is a free scale and thus $\sigma/\mu$ is the only parameter of this toy model.

\begin{figure}[t]
    \centering
    \includegraphics[width=\textwidth]{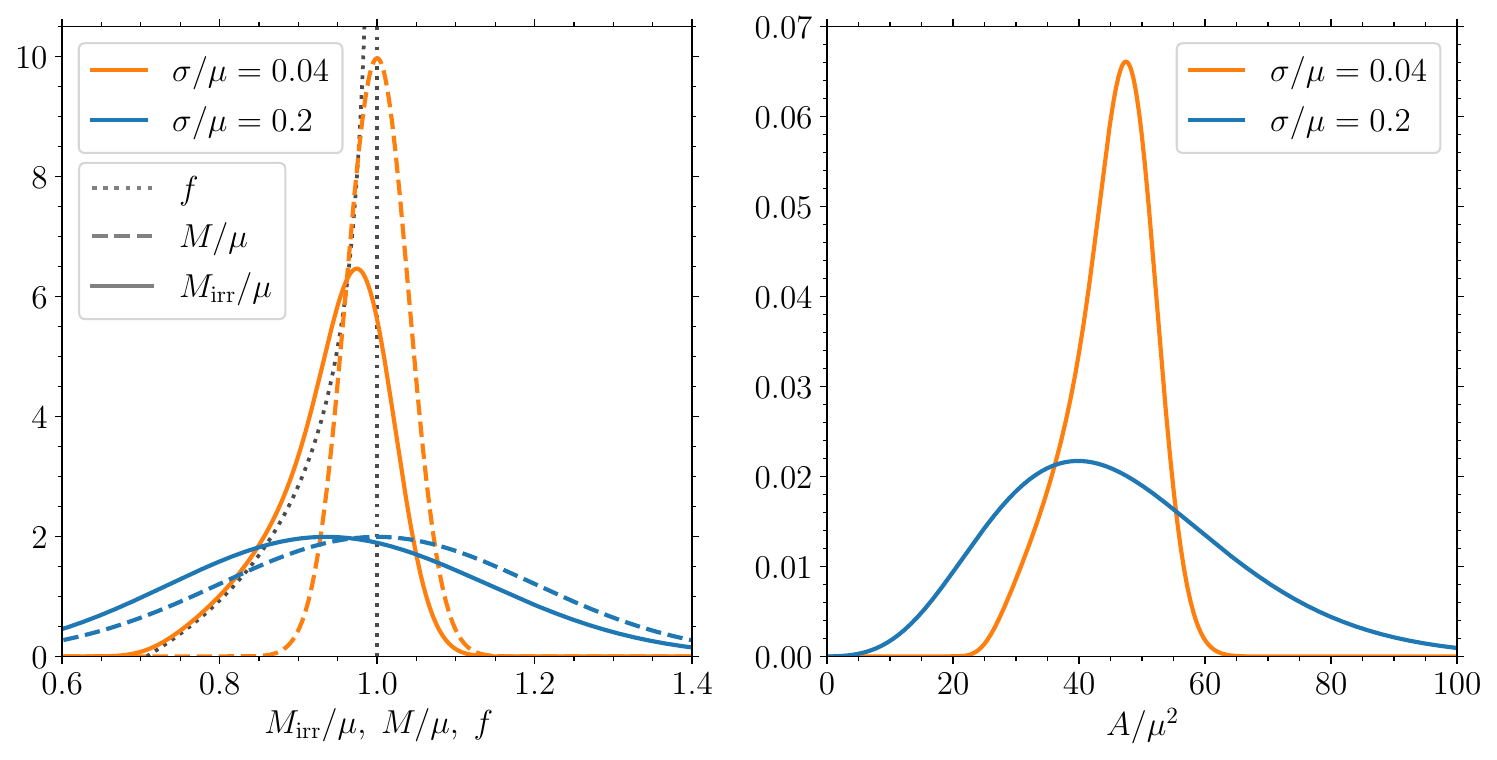}
    \caption{Probability density functions of the irreducible mass $M_{\rm irr}$ (left panel, solid), the BH mass $M$ (left panel, dashed), their ratio $f$ (left panel, dotted), and horizon area $A$ (right panel, solid) as obtained for the simple model of Sec.~\ref{simplemodel}. Quantities are expressed in units of the mean value of the BH mass $\mu$. We show distributions for two different values of the width parameter $\sigma/\mu = 0.04$  (orange) and $0.2$ (blue). }
    \label{pdfs}
\end{figure}

The left panel of Fig.~\ref{pdfs} highlights the distributions of  $f$, $M$ and $\Mirr$ for two representative values of $\sigma$. As $\sigma$ increases, the irreducible mass $\Mirr$ transitions from being distributed like $f$ to being distributed like $M$. In symbols, this corresponds to the following notable limits of Eq.~(\ref{pmirr}): %
\begin{alignat}{10}
p\left(\Mirr\right)  &\simeq \frac{\pi(\Mirr/\mu)}{\mu} \qquad&&{\rm for}\quad {\sigma}\ll \mu\,, \label{limit1}\\
p(\Mirr)  &\simeq \mathcal{N}(\mu,\sigma) \qquad && {\rm for}\quad \sigma\gg \mu \,. \label{limit2}
\end{alignat}
Intuitively, this means that large uncertainties of the BH mass will dominate our inference on $\Mirr$. If instead the BH mass is well measured, our uncertainty on $\Mirr$ will be dominated by the spin measurement (or lack thereof, as assumed in this toy model).

One can better quantify the properties of $p(\Mirr)$ by performing a Kolmogorov-Smirnov (KS) test, which is a standard frequentist tool to asses if two samples share the same probability distribution. The  KS distance between two samples $a$ and $b$ is defined as 
the largest absolute difference between the two empirical cumulative distribution functions \cite{2019sdmm.book.....I}.
Small (large) values of the statistics ${\rm KS}(a,b)$ correspond to large (small) $p$-values,
indicating that we cannot (can) reject the null hypothesis that the samples $a$ and $b$ were drawn from the same distribution.

\begin{figure}[t]
    \centering
    \includegraphics[height=0.6\columnwidth]{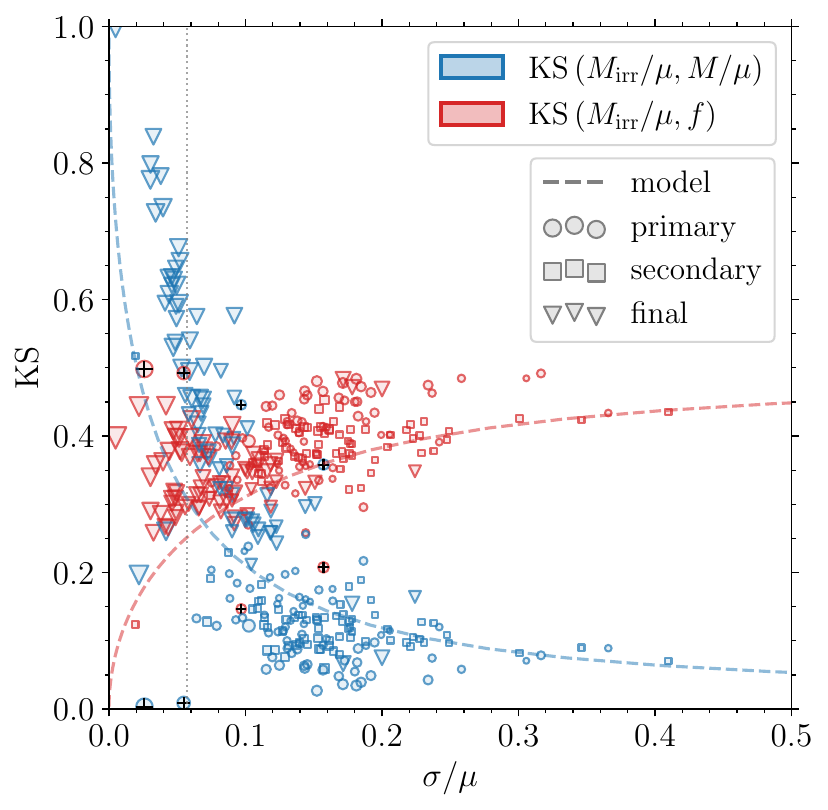}
    \caption{Kolmogorov-Smirnov (KS) distances between the irreducible mass $\Mirr$ and either the BH mass $M$ (blue) or the spin factor $f$ (red). The parameters $\mu$ and $\sigma$ are the median and the width of the mass distribution. Dashed curves are computed using the toy model of Sec~\ref{simplemodel}.
   The vertical dotted line at $\sigma/\mu= 0.06$ indicates an estimate of the transition value obtained by matching the 90\% credible intervals of the two limiting distributions, cf. Eqs.~(\ref{limit1}-\ref{limit2}). 
     Scatter points indicate the sample of 220 individual BHs measured by LIGO/Virgo. Circles, squares, and triangles mark primary components, secondary components, and post-merger remnants, respectively. The marker size indicates the accuracy of the spin measurement, with smaller (larger) %
    markers corresponding to wider (narrower) spin magnitude posteriors. %
     Some notable outliers among the component BHs are indicated with black crosses, see Sec.~\ref{individual}. }
    \label{KSevents}
\end{figure}

Figure~\ref{KSevents} compares the distribution of $M_{\rm irr}$ to those of $f$ and $M$. Results from our toy model are shown with dashed curves. When $\sigma/\mu\to 0 $,  the vast majority of the $M$ probability is concentrated over a region that is much smaller than the  support of the $f$ distribution. This implies ${\rm KS}(M_{\rm irr}/\mu , f)\to 0$, which corresponds to the limit in Eq.~(\ref{limit1}).  Conversely, for $\sigma/\mu\to \infty$, the support of the distribution of $M$ is much larger than that of $f$ and thus  ${\rm KS}(M_{\rm irr}/\mu , M/\mu)\to 0$, 
c.f. Eq.~(\ref{limit2}). A rough estimate of the transition value between these two behaviors can be estimated by matching the 90\% credible intervals of the limiting distributions $\pi(f)/\mu$ and $\mathcal{N}(\mu,\sigma)$. This yields $\sigma\simeq 0.06 \mu$
 as shown with the vertical dotted line in Fig.~\ref{KSevents}.   The remaining limits can also be easily understood. One has $\lim_{\sigma/\mu\to 0}{\rm KS}(M_{\rm irr}/\mu , M/\mu) \simeq  \lim_{\sigma/\mu\to 0}{\rm KS}(f , M/\mu)\to 1$ which correspond to the KS distance between distributions with non-overlapping supports. Similarly, we find $\lim_{\sigma/\mu\to \infty}{\rm KS}(M_{\rm irr}/\mu , f) \simeq  \lim_{\sigma/\mu\to \infty}{\rm KS}(M/\mu,f)\to 1/2$, which corresponds to the limit where the support of one distribution is infinitely larger than that of the other.
 
This calculation can be generalized to obtain the probability density function of the horizon area $A =16\pi M^2 f^2$. The result is more convoluted but nonetheless relatively compact: 
\begin{equation}
p(A)  = \frac{1}{2^{3/2} \pi \sigma \sqrt{A}} \int_{1/\sqrt{2}}^1 \exp\left[ - \frac{A/(16 \pi f^2) + \mu^2}{2\sigma^2}\right]   \cosh\left(\frac{\sqrt{A}\mu}{4 \sqrt{\pi} f \sigma^2}\right) \frac{2 f^2-1}{f \sqrt{1 - f^2}}  \,\, {\rm d}f\,. \label{pA}
\end{equation}
As before, this expression is valid if the spin magnitude $\chi$ is uniformly distributed in $[0,1]$ and the mass $M$ follows a Gaussian distribution with mean $\mu$ and variance $\sigma$. The resulting distributions are shown in the right panel of Fig.~\ref{pdfs}.

\section{Individual gravitational-wave events}
\label{individual}

We now turn our attention to the BHs observed by LIGO and Virgo. We consider the 76 compact-binary mergers listed in Table 1 of Ref.~\cite{2021arXiv211103634T}, which were selected from a longer list of triggers using a false-alarm rate threshold of $<1 ~{\rm yr}^{-1}$. We consider three objects for each event: the primary (heavier) component, the secondary (lighter) component, and the post-merger remnant. We then select BHs and discard neutron stars setting a threshold of  $2.5 M_\odot$ to the median of the mass distribution. This results in a final sample  of 220 individual (though not independent!) BHs. We use posterior distributions from Refs.~\cite{2021PhRvX..11b1053A, 2021arXiv211103606T,2020MNRAS.499.3295R, 2021arXiv210801045T}, selecting the analyses therein that employ a distance prior that is uniform in comoving volume and source-frame time. Remnant masses and spins are computed using numerical-relativity fits \cite{2012ApJ...758...63B,2016ApJ...825L..19H,2016PhRvD..93l4066G}.

\begin{figure}[p]
    \centering
    \includegraphics[page=1,height=0.46\columnwidth]{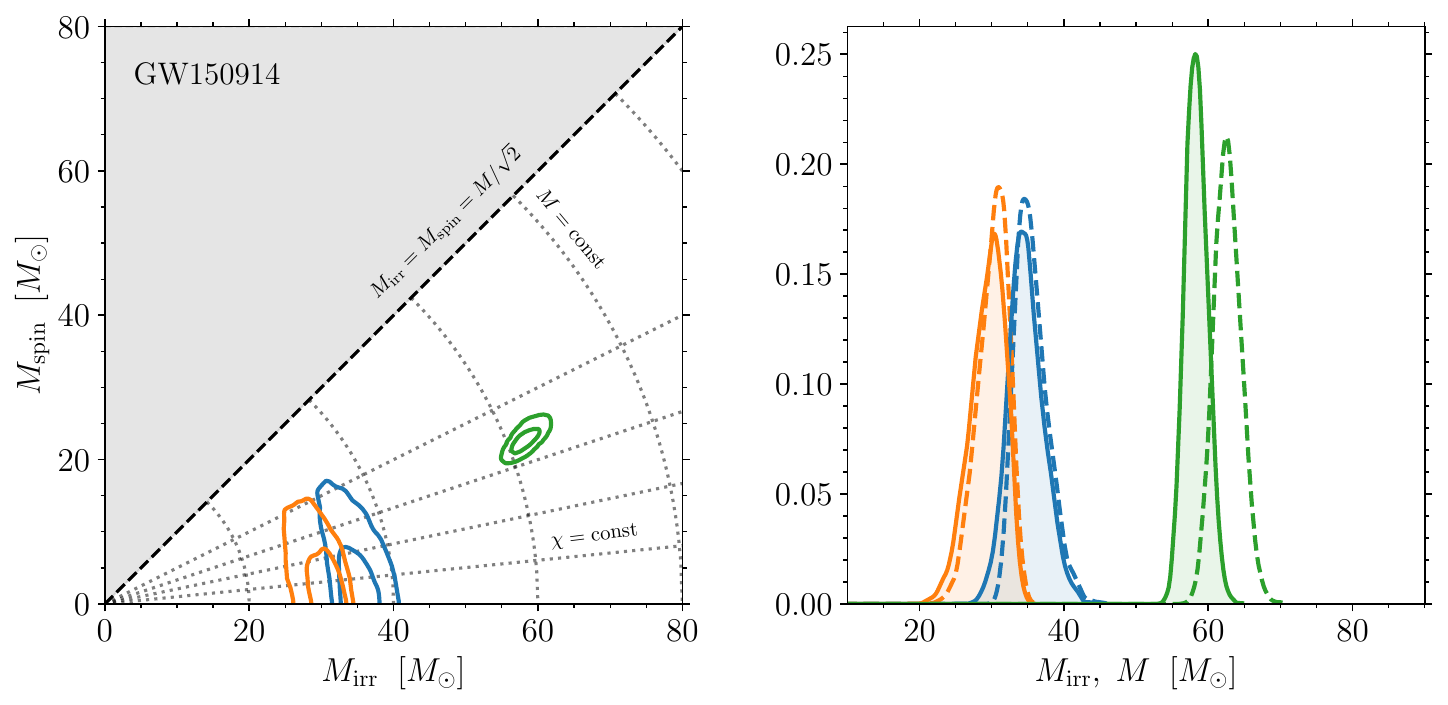}
        \includegraphics[page=2,height=0.46\columnwidth]{someKDEs}
    \includegraphics[page=3,height=0.46\columnwidth]{someKDEs}
    \caption{
 Distributions of the irreducible mass $\Mirr$, the rotational mass $\Mspin$, and the mass $M$ of the BHs for three representative GW events: GW150914 (top row), GW190517\_055101 (middle row), and GW191109\_010717 (bottom row). In both panels, blue, orange, and green curves indicate the heavier binary component, the lighter binary component, and the post-merger remnant, respectively. Panels in the left column show the joint distribution of $\Mirr$ and $\Mspin$ with contours indicating the 50\% and 90\% credible regions. Dotted circular curves correspond to the level sets $M/M_\odot= 20, 40, 60, \dots$; dotted radial lines correspond to $\chi = 0.2, 0.4,0.6,0.8 $. The black dashed line corresponds to $\chi=1$ and bounds the allowed region $\Mirr\geq \Mspin$. Panels in the right column show the marginalized distribution of $\Mirr$ (solid, filled) and $M$ (dashed, empty).
 }
    \label{someKDEs}
\end{figure}

Figure~\ref{someKDEs} illustrates the irreducible and rotational contributions to the masses of some of the observed BHs. 
The left panels show the joint posterior distribution of $\Mirr$ and $\Mspin$: this is a useful reparametrization of the BH mass-spin plane which manifestly highlights the two components of the total energy.  From Eqs.~(\ref{mirrdef}-\ref{mspindef}), one has $\Mspin\leq \Mirr$, and thus the upper region of this plot is forbidden. The diagonal line corresponds to $\chi=1$ and thus $\Mirr=\Mspin=M/\sqrt{2}$. Because $\Mirr$ and $\Mspin$ add quadratically to give the BH mass according to Eq.~(\ref{mtotcontr}), contours of constant $M$ are given by %
sectors of circles centered on the origin. 
On the other hand, the BH spin $\chi$ is constant along radial lines. The right panels of Fig.~\ref{someKDEs} show marginalized distributions of $\Mirr$ and $M$, in analogy with those illustrated earlier in Fig.~\ref{pdfs}.

For illustration, we discuss three GW events in more detail:
\begin{itemize}
\item The top panels of Fig~\ref{someKDEs} refer to the first GW observation, GW150914 \cite{2016PhRvL.116f1102A}. This is a system where the masses of the merging objects are relatively well measured but the spins are not, which is representative of the majority of the detected events to date. For the primary and secondary BHs, the 2-dimensional  $\Mirr-\Mspin$ distributions approximately follow circles of constant $M$ and only exclude the  region with $\chi\gtrsim 0.8$. Both BHs are compatible with having small spins and thus $\Mirr\lesssim M$ (right panel). In contrast, the post-merger BH has a well estimated spin of $\ssim 0.7$: the contours therefore extend approximately radially  in the ($\Mirr-\Mspin$) plane while the marginalized distribution of $\Mirr$ differs 
visibly
from that of~$M$.

\item The middle panels of Fig~\ref{someKDEs} illustrate GW190517\_055101, which is an event where the spin of the primary BH is relatively well measured \cite{2021PhRvX..11b1053A}. This translates into contours that exclude $\Mspin=0$ with high confidence: the primary BH of GW190517\_055101 definitely  had a fraction of its mass that one could theoretically extract via Penrose processes. The accuracies on $M$ and $\chi$ for the primary BH are such that $\Mirr$ and $\Mspin$ are approximately uncorrelated (i.e. their iso-probability contours are close to circular).

\item The bottom panels of Fig.~\ref{someKDEs} show results for GW191109\_010717 \cite{2021arXiv211103606T}. This is another of the few events where the rotational mass of the primary BH can be  constrained away from zero.  In this case, however, the spin of the post-merger BH is $\chi_{\rm f}\gtrsim 0.4$ ---a somewhat unusual occurrence for comparable-mass BH mergers~\cite{2021ApJ...915...56G}. Because of this, the irreducible mass of the remnant normalized to the total mass of the binary is larger than that of most of the other events (cf. Sec.~\ref{increasearea}).

\end{itemize}

Building on the investigation presented in Sec.~\ref{simplemodel}, we compute  KS distances between $\Mirr$, $M$, and $f$ for all BHs in our  sample. The result is shown with scatter points in  Fig.~\ref{KSevents}, where in this context we take $\mu$ and $\sigma$ to be the median and the interquantile range of the mass distribution, respectively (the latter is normalized such that it becomes an unbiased estimator of the standard deviation for Gaussian data, see e.g. Ref.~\cite{2019sdmm.book.....I} for details). 

 Our toy model from Sec.~\ref{simplemodel} (dashed lines) works relatively well for many of the pre-merger BHs (circles and squares in Fig.~\ref{KSevents}).  
Recall that the key assumption behind that model is that the spin is poorly measured. The measured values and the semi-analytic predictions are expected to differ significantly for objects with  well measured spins. The outliers seen in the subsamples of pre-merger BHs from Fig.~\ref{KSevents} (black crosses) are the primaries of GW190814, GW190517\_055101, GW191109\_010717,  GW200105\_162426 which are, indeed, all systems where the spin could be  meaningfully constrained  \cite{2021PhRvX..11b1053A,2021arXiv211103606T}.
The subsample of final BHs (triangles in Fig.~\ref{KSevents}) is not well described by our semi-analytical model. The spins of BH remnants are  typically close to $\chi \sim 0.7$ (almost independently of the parameters of the merging binary \cite{2017PhRvD..95l4046G,2017ApJ...840L..24F,2021ApJ...915...56G,2021CQGra..38d5012G}), thus violating the assumptions behind our semi-analytic model. In particular, we find that both distances ${\rm KS}(\Mirr/\mu, M/\mu)$ and ${\rm KS}(\Mirr/\mu, f)$ 
are preferentially larger than those we predicted in Sec.~\ref{simplemodel}. In general, this indicates that the distribution of the irreducible mass of post-merger BH remnants cannot be approximated by any of the limits reported in  Eqs.~(\ref{limit1}-\ref{limit2}).

\begin{figure}[p]
    \centering
    \includegraphics[width=\columnwidth]{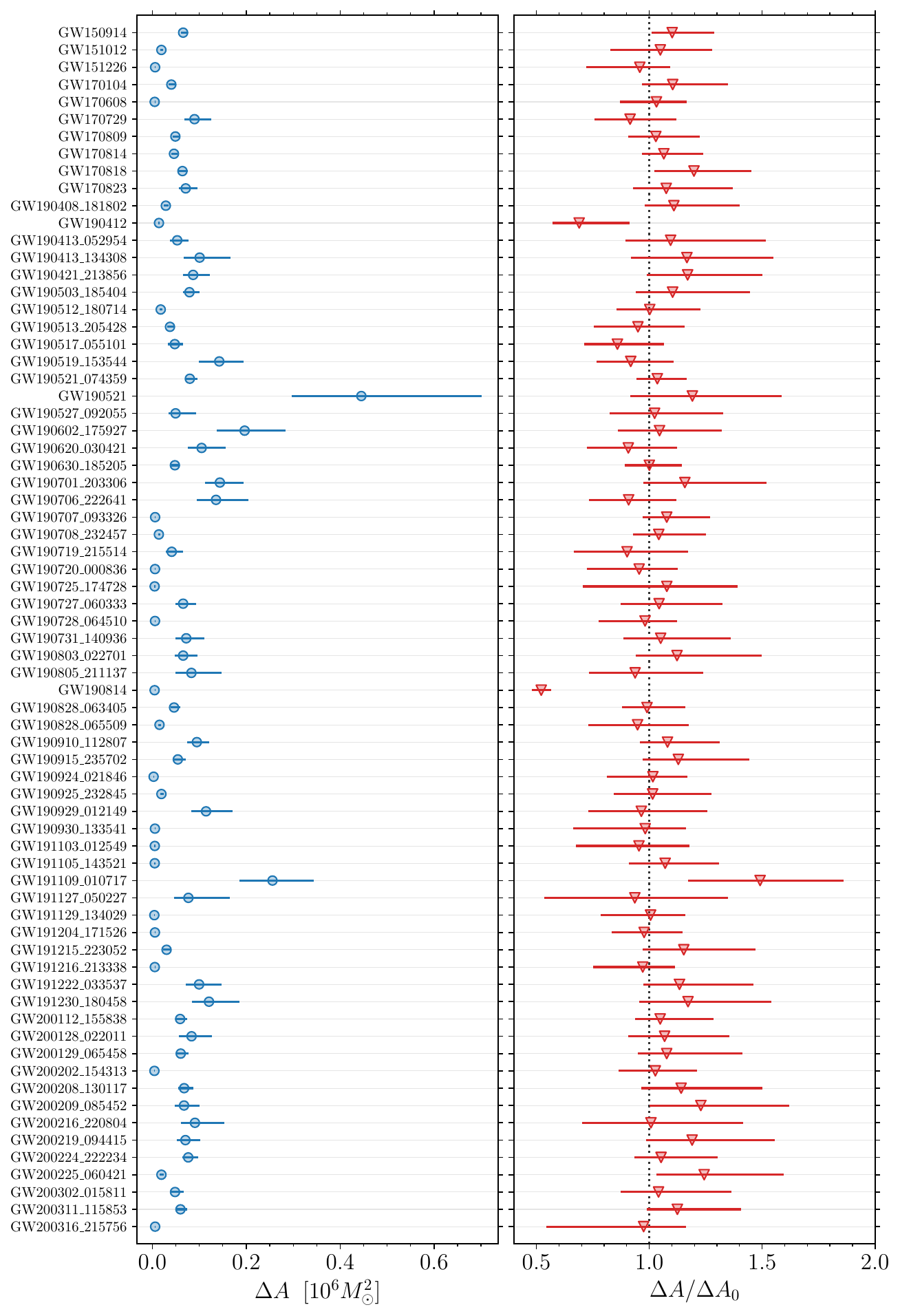}
    \caption{Increase of the total horizon area $\Delta A$ for the merging binary BHs detected by LIGO/Virgo. The left panel (blue circles) indicates $\Delta A$ in units of $M_\odot^2$ with $c=G=1$; the right panel (red triangles) indicates $\Delta A$ normalized to the value $\Delta A_0\simeq 13.75 M_{\rm tot}^2$ one naively expects for a BH merger (dotted line),  see~Sec.~\ref{increasearea}. Markers indicate the median values of the posterior distributions; error bars correspond to the 90\% credible intervals. Events are reported in chronological order from top to bottom.}
    \label{deltAall}
\end{figure}

\section{Increase of horizon area}
\label{increasearea}

We can also  quantify the increase of horizon area (i.e. entropy) that occurred when these BHs merged. We compute  the variation 
\begin{equation}
\Delta A \equiv A_{\rm f} - A_1 - A_2 = 
16 \pi (M_{\rm irr, f}^2 - M_{\rm irr, 1}^2 - M_{\rm irr, 2}^2)\,,
\end{equation} where the subscripts $1$, $2$, and ${\rm f}$ refer to the primary, secondary and post-merger BHs, respectively. Hawking's area law states that $\Delta A\geq 0$ \cite{1971PhRvL..26.1344H}. Figure~\ref{deltAall} shows the resulting constraints for those 70 GW events where both the primary and the secondary are BHs (thus discarding all systems with at least one neutron star).  Note that this is not a test of the area law but rather a quantification  because the underlying data were analyzed assuming that General Relativity (and thus the area law) holds true. 

For comparison, one can estimate the entropy variation $\Delta A_0$ assuming a merger of two equal-mass, non-spinning BHs. In this case one has $M_{\rm irr,1} = M_{\rm irr,2}= M_1=M_2 = M_{\rm tot}/2$, where $M_{\rm tot}=M_1+M_2$ is the binary total mass. This yields $\Delta A_0 = 16 \pi (M_{\rm irr, f}^2 - M_{\rm tot}^2/ 2)$. The final BH has irreducible mass $  M_{\rm irr, f} = (1-\lambda) M_{\rm tot}  \sqrt{\left(1+\sqrt{1-\chi_f^2} \right)/2}$ where $\lambda$ is the energy fraction dissipated during the merger process and $\chi_{\rm f}$ is the post-merger BH spin. Early numerical-relativity simulations~\cite{2005PhRvL..95l1101P,2009PhRvD..79b4003S} showed that  $\lambda \sim 5\%$ and $\chi_{\rm f}\sim 0.7$,  which results in $\Delta A_0\simeq 13.75 M_{\rm tot}^2$. 

As shown in Fig.~\ref{deltAall}, we find that the horizon-area increase  inferred from the data is compatible with this simple back-of-the-envelope estimate for most of the observed events.
The two notable exceptions with $\Delta A\lesssim \Delta A_0$ are GW190412 and GW190814: these are systems with a mass ratio that significantly departs from unity and thus one would expect $A_{\rm f} \gtrsim A_1$. The event with the largest absolute value of $\Delta A$ is GW190521 (left panel of Fig.~\ref{deltAall}), which is likely to be the most massive event in the sample. Less trivially, the event that shows the largest area change  relative to its total mass  is  GW191109\_010717 (right panel of Fig.~\ref{deltAall}). Quoting median and 90\% credible interval, we find $\Delta A/\Delta A_0 = 1.49^{+0.32}_{-0.37}$. The properties of this event are highlighted in the bottom panels of Fig.~\ref{someKDEs}. GW191109\_010717  can be regarded as the ``most irreversible'' BH merger in the current GW catalog. The ranking reported in Fig.~\ref{deltAall} provide a useful guide to selecting events for test of GR based on the area law \cite{2018PhRvD..97l4069C,2021PhRvL.127a1103I}, as a higher degree of irreversibility might provide more constraining power.

\section{Constraints from the inferred population}
\label{popsec}

We now turn our attention to the following question: given the entirety of the LIGO data collected so far, what is the rate of energy that the Universe inexorably stores  inside BH horizons? What is the rate of entropy increase implied by current data?

To tackle this point one needs to go beyond single-event inference and consider population fits.   Let us denote the parameters of the individual GW events with $\theta$  (e.g. masses $m_{1,2}$, spins $\boldsymbol\chi_{1,2}$, redshift $z$)  and the parameters of the  population with $\lambda$  (e.g. local merger rate, spectral index of the mass spectrum, etc.). We use the default model of Ref.~\cite{2021arXiv211103634T} as implemented in Ref.~\cite{2019PhRvD.100d3030T}. In particular, the differential merger rate ${{\rm d} \mathcal{R} }/{{\rm d} \theta}$ is assumed to be proportional to a distribution  $p_{\rm pop} $ of masses and spins and a redshift-dependent term  $R$ that encodes the normalization. In symbols, this is
\begin{equation}
\frac{d \mathcal{R}}{d\theta}(\theta | \lambda) = {R}(z|\lambda_R) \times p_{\rm pop}  (m_1,m_2,\boldsymbol\chi_1,\boldsymbol\chi_2| \lambda_{\rm pop})\,,
\label{drdthetafactor}
\end{equation}
where $\theta=\{m_1,m_2,\boldsymbol\chi_1,\boldsymbol\chi_2, z\}$, $\lambda = \{\lambda_R,\lambda_{\rm pop}\}$, and $\int p_{\rm pop} d\theta=1$. The probability $p_{\rm pop}$ is constructed such that   $m_1$ is distributed according to a truncated power-law with an additional Gaussian component, $m_2$ conditioned on $m_1$ %
is distributed according to a power-law, the spin magnitudes $\chi_{1,2}$ are modeled with a beta distribution, and the spin directions $\theta_{1,2}$ are modeled as superposition of an isotropic component and a truncated Gaussian (cf. Ref.~\cite{2021arXiv211103634T} and references therein). The normalization is set to $R(z| R_0,\kappa)= R_0 (1+z)^\kappa$ where $R_0$ is the merger rate in the local Universe ---a quantity that is typically measured in Gpc$^{-3}$yr$^{-1}$. In particular, $\kappa=0$ corresponds to a merger rate that is uniform in comoving volume and source-frame time. With the data at hand, we can only model sources that lie within the LIGO horizon redshift, i.e. $z \lesssim z_{\rm H}$. We take $z_{\rm H}=1.5$, which is estimated from Fig. 13 in Ref.~\cite{2021arXiv211103634T}. It is understood that a rate parametrization $\propto (1+z)^\kappa$ cannot be valid at arbitrarily large redshifts because the star formation rate decreases substantially for $z\gtrsim 2$ \cite{2014ARA&A..52..415M}.
Using a hierarchical Bayesian scheme, Ref.~\cite{2021arXiv211103634T} fitted the phenomenological model we just described to the available GW events  and released samples from the posterior distribution $p(\lambda)$.

\begin{figure}[t]
    \centering
    \includegraphics[height=0.6\columnwidth]{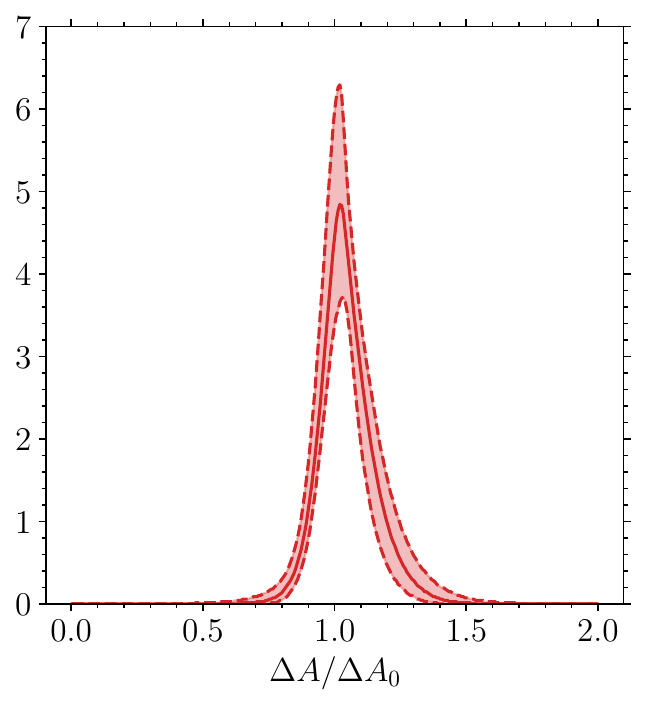}
    \caption{Distribution of relative entropy increase per merger $\Delta A/A_0$ implied by current LIGO/Virgo data. Solid and dashed curves indicate the median and the 90\% credibile interval of the distribution, respectively. The scaling $\Delta A_0\simeq 13.75 M_{\rm tot}^2$ is obtained from a simple, back-of-the-envelope estimate; see~Sec.~\ref{increasearea}.}
    \label{poppostdist}
\end{figure}

\begin{figure}[t]
    \centering
    \includegraphics[width=\columnwidth]{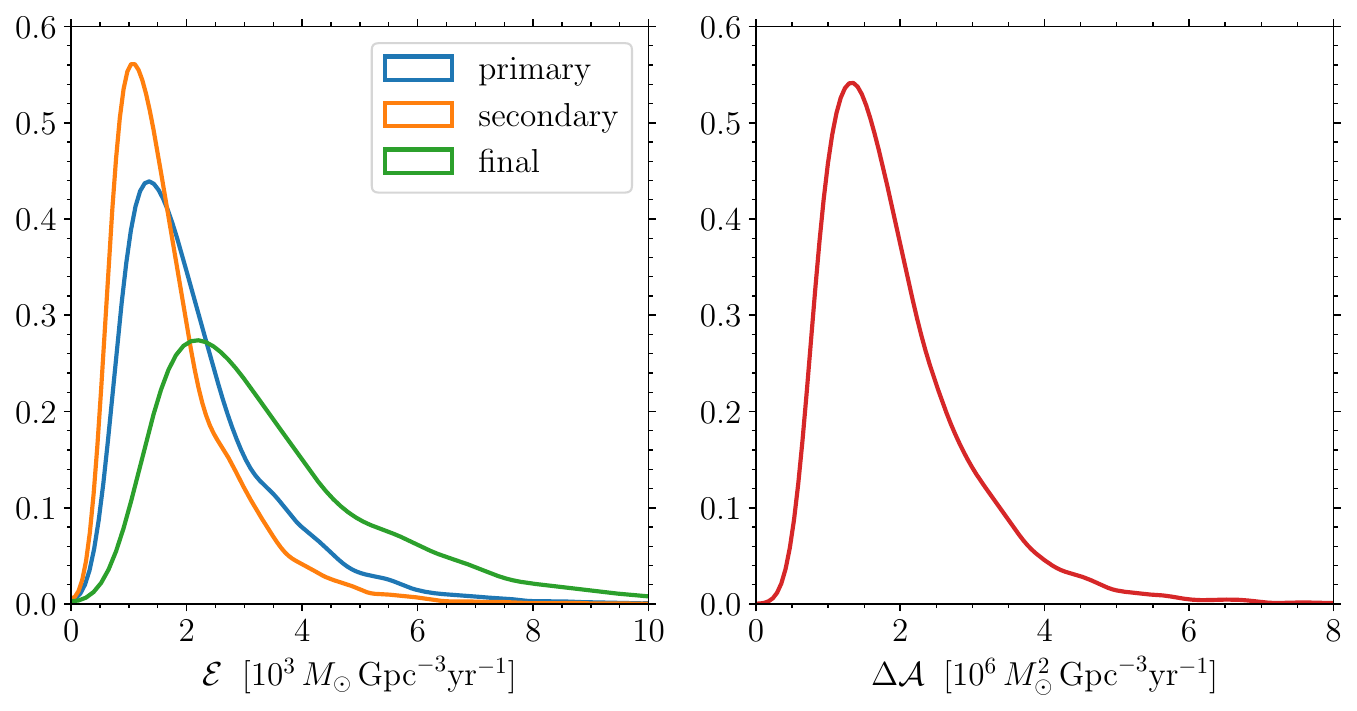}
    \caption{Left panel: Rate of energy that is forever stored inside BH horizons, as inferred from current LIGO/Virgo data. The quantity $\mathcal{E}$ has dimension of energy over time over volume [Eq.~(\ref{bigEi})]. Blue, orange, and green curves refer to primary BHs, secondary BHs, and post-merger remnants, respectively. Right panel: Variation of horizon area per unit volume and time due to BH mergers detectable by LIGO/Virgo. The quantity $\mathcal{A}$ has dimension of area over time over volume~[Eq.~(\ref{bigAi})]. }     \label{rateaveraged}
\end{figure}

First, we compute the distribution of entropy increase per merger. For each population sample $\lambda_i\sim p(\lambda)$, we extract a mock population $\theta_{ij} \sim p_{\rm pop}(\theta | \lambda_j)$, evaluate $\Delta A(\theta_{ij})$, and estimate $p(\Delta A| \lambda_j)$. Figure~\ref{poppostdist} shows the resulting median and $90\%$ credible intervals. Once more, we normalize $\Delta A/A_0$ using our back-of-the-envelope estimate $A_0\simeq 13.75 M_{\rm tot}^2$ from Sec.~\ref{increasearea}. The distribution of relative entropy increase $\Delta A/ A_0$ is strongly peaked around $\ssim 1$ with vanishing tails $\lesssim 1/2$ and $\gtrsim 2$. From the median distribution over $\lambda$, we report $p(\Delta A/A_0 > 1.2)\simeq 8\% $ and $p(\Delta A/A_0 > 1.4) \simeq 0.4\%$. In this context, the event GW191109\_010717 highlighted above ($\Delta A/\Delta A_0 = 1.49^{+0.32}_{-0.37}$) should be regarded as a moderately extreme outlier. This is worthy of a more detailed investigation as, to the best of our knowledge, none of the GW catalog analyses performed so far highlighted GW191109\_010717 as a catalog outlier.

The energy per unit comoving volume and unit time that one can never hope to extract from BHs is given by
\begin{equation}
\mathcal{E}_i (\lambda) = c^2 \int M_{{\rm irr}}(m_i,\chi_i)  \; \frac{d \mathcal{R}}{d\theta}(\theta | \lambda) \,d\theta \,,
\label{bigEi}
\end{equation}
where we reinstated a factor of $c^2$ to stress that the quantity $\mathcal{E}_i$ has the dimension of energy over space-time volume. With the same notation used above, the index $i$ labels either the primary BHs ($i=1$), the secondary BHs ($i=2$), or the post-merger BH remnants ($i={\rm f}$). Because of the factorization of Eq.~(\ref{drdthetafactor}), the redshift integral in Eq.~(\ref{bigEi}) can be carried out analytically, resulting in an overall multiplicative factor\footnote{The value of $r_{\rm H}$ enters our results explicitly because we are restricting our calculation to the portion of the Universe that is accessible to LIGO.} equal to $R_0 [{(z_{\rm H}+1)^{\kappa+1} - 1}]/({\kappa+1})$. The remaining integrals in the mass and spin dimensions can be approximated as Monte Carlo summations using samples drawn from $p_{\rm pop}$.  
The resulting Bayesian measurements of $\mathcal{E}_i$ are shown in the left panel of Fig.~\ref{rateaveraged}. %
We find %
$\mathcal{E}_1= 1.86^{+3.03}_{-1.07} \times 10^{3} M_\odot {\rm Gpc}^{-3} {\rm yr}^{-1}$, 
 $\mathcal{E}_2= 1.47^{+2.40}_{-0.83} \times 10^{3} M_\odot {\rm Gpc}^{-3} {\rm yr}^{-1}$, and
  $\mathcal{E}_{\rm f}= 3.00^{+4.90}_{-1.72} \times 10^{3} M_\odot {\rm Gpc}^{-3} {\rm yr}^{-1}$.
 Note that  $\mathcal{E}_{\rm f}\lesssim \mathcal{E}_{1} + \mathcal{E}_{2}$ but this is not in contradiction with the area law because $A\propto M_{{\rm irr}}^2$.

Similarly, one can compute the horizon area of BHs within the LIGO range per unit volume and unit time:
\begin{equation}
\mathcal{A}_i (\lambda) = \frac{G^2}{c^4} \int A(m_i,\chi_i)  \; \frac{d \mathcal{R}}{d\theta}(\theta | \lambda) \,d\theta \,,
\label{bigAi}
\end{equation}
where again we reinstated the fundamental constants for clarity. Using current LIGO data we find $\mathcal{A}_1= 1.89^{+2.60}_{-1.00} \times 10^{6} M_\odot^2 {\rm Gpc}^{-3} {\rm yr}^{-1}$, $\mathcal{A}_2= 1.11^{+1.42}_{-0.56} \times 10^{6} M_\odot^2 {\rm Gpc}^{-3} {\rm yr}^{-1}$, and $\mathcal{A}_{\rm f}= 4.69^{+6.12}_{-2.45} \times 10^{6} M_\odot^2 {\rm Gpc}^{-3} {\rm yr}^{-1}$. The rate of horizon-area increase $\Delta{\mathcal{A}}= \mathcal{A}_{\rm f} - \mathcal{A}_1 -\mathcal{A}_2$ is shown in the right panel of Fig.~\ref{rateaveraged}. Just considering systems that are targeted by LIGO, the total area of BH horizons increases by $1.69^{+2.19}_{-0.88} \times 10^{6} M_\odot^2$ each year in each ${\rm Gpc}$ cubed of the Universe (at least up to $z_{\rm H}$). We stress that this is an extremely conservative lower limit to the total BH-area increase of the Universe because it only takes into account systems that emit GWs at frequencies within the LIGO/Virgo sensitivity window ($\sim 10-1000$ Hz). That said, it is important to stress that the population fits of Ref.~\cite{2021arXiv211103634T}  used in this analysis take selection effects into account: our estimate thus refers to BHs that are \emph{observable} by LIGO, not only to the specific catalog realization that has been \emph{observed}.

We stress that the calculation presented here assumes that GR holds true and thus $\Delta{\mathcal{A}}\geq 0$ 
 by construction. Our quantification provides the baseline value for future tests that combine multiple GW events statistically \cite{2019PhRvD..99l4044Z,2019PhRvL.123l1101I,2021iSci...24j2577M}. Testing the area law within this framework is a promising yet unexplored avenue for future research.

\section{A useful reparametrization}
\label{concl}

Building on the concept of irreducible mass~\cite{1970PhRvL..25.1596C} and horizon area \cite{1971PhRvL..26.1344H}, we have presented a new exploitation of current LIGO/Virgo data, studying in particular (i) the energy that is stored inside BH horizons and (ii) the increase of entropy.

We have developed a simple model of the expected posterior distributions (Sec.~\ref{simplemodel}) and used it as guidance to analyze individual GW events (Sec.~\ref{individual}). In a nutshell, our explorations can be summarized in the following reparametrization of the BH properties:
\begin{equation}
(M,\chi) \longleftrightarrow (\Mirr,\Mspin)\,
\end{equation}
While a Kerr BH is usually described by its mass $M$ and spin $\chi$, its energy contributions are best expressed in terms of the irreducible mass $\Mirr$ and the rotational mass $M_{\rm spin}$. The explicit transformations are given by
\begin{equation}
\begin{cases}
\displaystyle
\Mirr =  M \sqrt{\frac{1+\sqrt{1-\chi^2}}{2}}\,, \\
\displaystyle
\Mspin =  M \sqrt{\frac{1-\sqrt{1-\chi^2}}{2}}\,,
\end{cases}
\qquad
\begin{cases}
\displaystyle
M =  \sqrt{\Mirr^2 +\Mspin^2}\,, \\[10pt]
\displaystyle
\chi =  \frac{2 \Mirr \Mspin}{(\Mirr^2 +\Mspin^2)}\,.
\end{cases}
\end{equation}
The application of this reparametrization to the data is made most evident in the left panels of Fig.~\ref{someKDEs}. %

Among the systems detected with false-alarm rate  $<1 ~{\rm yr}^{-1}$, we find that the BH merger that caused the largest entropy variation relative to the  total mass of the binary is GW191109\_010717. This event is thus a promising target for tests of relativity searching for  possible violations of Hawking's area law. Looking at the entire set of detected systems (Sec.~\ref{popsec}), this event appears to be an outlier of the detected population ---a feature of the catalog that is worthy of a more detailed investigation.

We have also formulated a ``population version'' of the area law (Sec.~\ref{popsec}). This allows us to quantify for the first time the rate of energy that is stored inside BH horizons as well as the overall rate of entropy increase implied by current data. We report $\Delta\mathcal{A} \sim 2 \times 10^{6} M_\odot^2  {\rm Gpc}^{-3} {\rm yr}^{-1}$ ($\sim 4\times 10^{-72} {\rm m}^{-1} {\rm s}^{-1}$ in SI units).

The separation of the BH mass into its  irreducible and rotational components is a fundamental property of the Kerr geometry with important consequences in theoretical physics. While this work is restricted to characterizing the available data, we hope it will spark further investigations aimed at constraining the irreversibility of BH mergers observationally with GWs.

\section*{Acknowledgments}

We thank Matthew Mould and Matteo Bonetti for discussions. We thank Maya Fishbach for sharing her code to sample $p_{\rm pop}$. 
D.G. and C.M.F. are supported by 
European Union's H2020 ERC Starting Grant No.~945155--GWmining, 
Cariplo Foundation Grant No.~2021-0555, 
and 
Leverhulme Trust Grant No.~RPG-2019-350. 
U.S.~is supported by
STFC Research Grant Nos. ST/V005669/1
and ST/W001667/1,
DiRAC project ACT284, funded by STFC grants
ST/P002307/1,
ST/R002452/1 and
ST/R00689X/1,
and
NSF-XSEDE Grant No.~PHY-090003.
Computational work was performed at CINECA with allocations through INFN, Bicocca, and ISCRA Type-B project HP10BEQ9JB, as well as at
SDSC Expanse, TACC Stampede, the Cambridge Service for Data
Driven Discovery (CSD3) system,
and the Maryland Advanced Research
Computing Center (MARCC).

\section*{ORCID iDs}
Davide Gerosa~~\orcidlink{0000-0002-0933-3579} \href{https://orcid.org/0000-0002-0933-3579}{https://orcid.org/0000-0002-0933-3579} \\
Cecilia Maria Fabbri~~\orcidlink{0000-0001-9453-4836} \href{https://orcid.org/0000-0001-9453-4836}{https://orcid.org/0000-0001-9453-4836}\\
Ulrich Sperhake~~\orcidlink{0000-0002-3134-7088} \href{https://orcid.org/0000-0002-3134-7088}{https://orcid.org/0000-0002-3134-7088} \\

\section*{References}
\bibliographystyle{iopart-num_leo}
\bibliography{irreduciblemass}

\end{document}